\documentclass{article}

\usepackage{cite,graphicx,subfigure,setspace}
\usepackage[small,bf,up]{caption}

\begin{document}
\title{Plasmonic Enhancement of Emission from Si-nanocrystals}
\author{Yiyang Gong, Jesse Lu, Szu-Lin Cheng, Yoshio Nishi, and Jelena Vu\v{c}kovi\'{c} \\
    \small\textit{Department of Electrical Engineering, Stanford University, Stanford, CA 94305}}
\twocolumn[
\begin{@twocolumnfalse}
\maketitle
\begin{abstract}
Plasmonic gratings of different periodicities are fabricated on top of Silicon nanocrystals embedded in Silicon Dioxide. Purcell enhancements of up to 2 were observed, which matches the value from simulations. Plasmonic enhancements are observed for the first three orders of the plasmonic modes, with the peak enhancement wavelength varying with the periodicity. Biharmonic gratings are also fabricated to extract the enhanced emission from the first order plasmonic mode, resulting in enhancements with quality factors of up to 16.
\end{abstract}
\end{@twocolumnfalse}
]
%\doublespacing
Enhancing the intensity and thus the emission rate of quantum emitters has garnered significant interest in the past decade. Research into such cavity quantum electrodynamics (cQED) effects has applications in the fields ranging from lasers to quantum information processing. In particular, the strength of cQED effects is increased as the quality ($Q$) factors of confined electromagnetic modes increase or their mode volumes decrease. The surface plasmon polariton (SPP), with ultra-small mode volume below the dielectric cavity limit of $(\lambda/(2n))^3$ (where $n$ is the index of refraction of the host material), has been suggested as one suitable candidate for cQED studies. In recent years, InGaN based quantum wells and CdSe colloidal quantum dots have been successfully coupled to the high density of optical states at the surface plasmon frequency \cite{Yablonovitch_SP, Scherer_SP,Koichi_SP}. In order to efficiently extract emission and to diversify the SPP enhancement over a range of wavelengths, several groups have investigated grating type structures combined with quantum wells or organic materials \cite{Shen_SPQW,Hung_SPdye,Chiu_SPorg}.

In this paper, we demonstrate plasmonic grating assisted enhancement of the emission from Silicon nanocrystals (Si-NCs) embedded in Silicon Dioxide in the 700-900nm wavelength range. This material is a viable option for building an inexpensive Si CMOS compatible light source for optical communications, interconnects, or solid-state lighting purposes \cite{Godefroo_SiNC,Pavesi_SiNC}. Due to the presence of metal, such a device can also be configured to be electrically injected \cite{Atwater_epump}. While such nanocrystals have been previously coupled to isolated metallic disks \cite{Biteen_SiNC}, coupling to gratings allows enhanced out-coupling and easier configurations for electrical injection.

\begin{figure}[ht]
\centering
\includegraphics[width=3.3in]{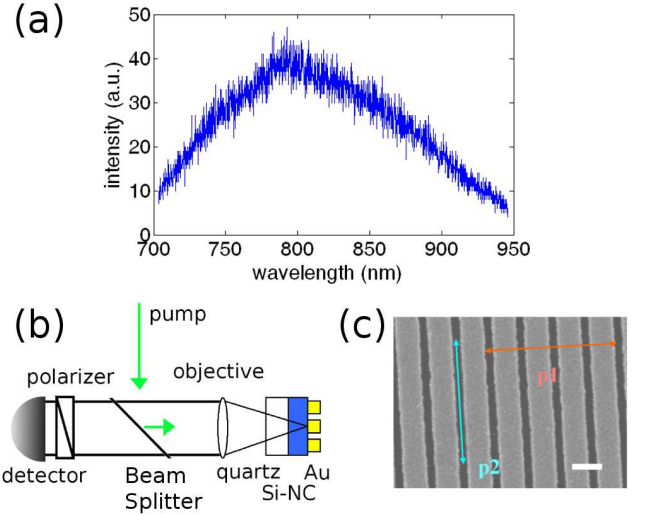}
\caption{(a) PL from the bulk Si-NC wafer (unprocessed, i.e., without metallic grating). (b) Experimental setup. (c) Fabricated gold grating, with p1 and p2 denoting the two polarizations selected in the experiments. Marker denotes 200nm.}
\label{fig:bulk}
\end{figure}

The studied structures are fabricated by the following procedure: A 70nm thick layer of Silicon Dioxide (SiO$_{2}$, $n=1.5$) is grown on polycrystalline quartz by plasma-enhanced chemical vapor deposition (PECVD) with a 14:1 mixture of SiH$_{4}$ (diluted to 2\% with N$_{2}$) and N$_{2}$O at 350$^\circ$C. The wafer is subsequently annealed in nitrogen at 1100$^\circ$C to form the Si-NCs. The Si-NC enriched SiO$_{2}$ layer has an effective refractive index of $n_{NC}=1.7$, and its bulk photoluminescence (PL) is shown in Figure \ref{fig:bulk}(a). Then, a 100nm layer of polymethylmethacrylate (PMMA) is spun on top of the wafer followed by subsequent e-beam lithography to define the grating pattern. Finally, a 30nm layer of gold (Au) is evaporated on top of the PMMA layer and the grating is fabricated by lift-off in acetone. The fabricated gratings have duty cyles of 0.6 to 0.8, but the overall trends of the PL data do not change over this range of duty cycles. Because the SPP mode of interest is confined to the Si-NCs-gold layer interface, we would expect the emission from the SPP mode to be redirected toward the Si-NCs and subsequently toward the quartz substrate. Hence, we pump Si-NCs with a green laser at 532nm, and collect the emission from the back side of the sample with an objective lens with large numerical aperture NA=0.5, maximizing the collection (Figure \ref{fig:bulk}(b)).

\begin{figure}[ht]
\centering
\includegraphics[width=3.3in]{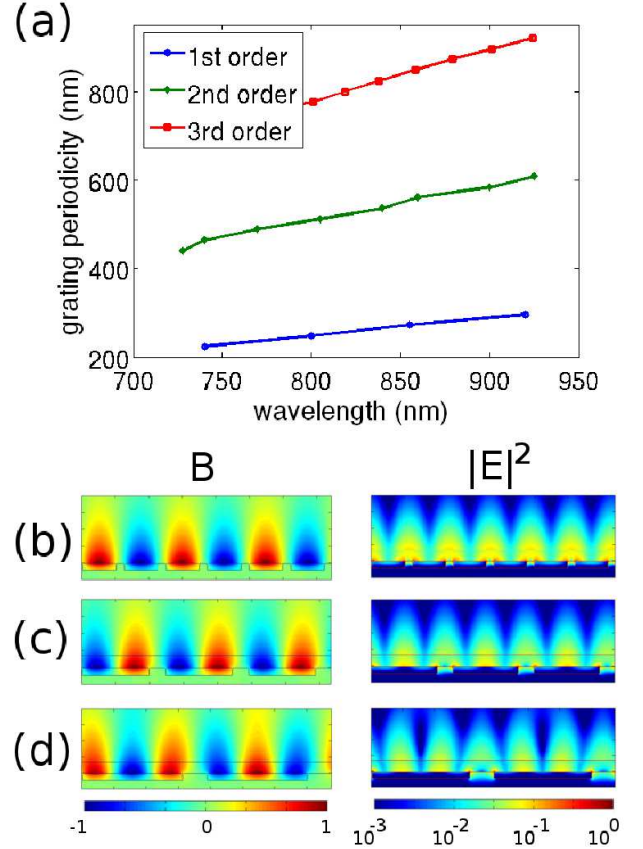}
\caption{(a) The FDTD calculated band edge frequencies for first, second, and third order modes of the SPP band diagram. The first and third order modes correspond to modes at the $X$ ($k=\pi/a$) point of the dispersion relation, while the second order mode corresponds to modes at the $\Gamma$ ($k=0$) point. The $|E|^2$ and $|B|^2$ fields for the first (b), second (c), and third (d) order modes are plotted.}
\label{fig:fdtd}
\end{figure}

We first analyze the exact structure in two dimensional (2D) finite difference time domain (FDTD) simulations as in previous references \cite{JV_sim, YG_sim}. We calculate the band edge frequencies of the modes at the $\Gamma$ ($k=0$) and $X$ ($k=\pi/a$) points of the band diagram, fixing a duty cycle of 0.7 and varying the grating period from 200-1000nm. We plot the free space mode wavelengths ($\lambda_{0}$) for first three relevant modes in the wavelength range of our PL in Figure \ref{fig:fdtd}(a), with the first order mode fitting one half SPP wavelength per grating period (corresponding to $k=\pi/a$, i.e. the $X$ point of the band diagram), the second order mode fitting a full SPP wavelength per grating period ($k=2\pi/a$, which folds to the $\Gamma$ point), and so forth. We also calculate the field profiles for the first three order modes for a duty cycle of 0.8, fixing the free-space wavelength $\lambda_{0}\approx 800$nm, and plotting the results in Figure \ref{fig:fdtd}(b)-(d). The grating period ($a_{p}$) for the p$^{th}$  order mode with a given mode frequency $\omega$ can be roughly determined from the SP dispersion relationship at the Si-NCs-gold interface without a grating \cite{Raether}:
\begin{equation}
    \frac{p\pi}{a_p}=k_{g}=\frac{\omega}{c}\sqrt{\frac{\epsilon_{d}\epsilon_{m}(\omega)}{\epsilon_{d}+\epsilon_{m}(\omega)}}
\label{dispersion_eq}
\end{equation}
where $\epsilon_{d}=n_{NC}^2$ is the dielectric constant of the Si-NC layer, the dielectric constant of gold is $\epsilon_{m}(\omega)=1-(\omega_{p}/\omega)^2$, and plasma frequency of gold is $\omega_{p}=2\pi c/(160\times10^{-9}\textrm{m})$. First, we calculate the Purcell enhancement by discretizing the dispersion relation by following the procedure from reference \cite{Hideo_SPtheory}. The Purcell enhancement for each k-vector is given by:
\begin{equation}
	F(\nu,k)=\frac{3}{2}\frac{c^3}{n_{NC}^3\omega^2(k)}\frac{\textrm{max}[\epsilon_{E}(z)|E(z)|^2]}{\int\epsilon_{E}(z)|E(z)|^2dz}\overline{\left(\frac{E}{E_{max}}\right)^2}D(\nu),
\end{equation}
where the normalized density of states is:
\begin{equation}
	D(\nu) = \frac{1}{\pi}\frac{\omega/2Q_{k}}{(\omega-\nu)^2+(\omega/2Q_{k})^2},
\end{equation}
$\omega(k)$ is the dispersion relation for the grating structure, $Q_{k}$ is the quality factor corresponding to the mode at the $(\nu,k)$ point of the band diagram, and $\epsilon_{E}(x,z)=d(\omega\epsilon(x,z))/d\omega$ is the effective dielectric constant. We then sum over all k-vectors to calculate the Purcell enhancement as a function of the frequency $\nu$. This enhancement is averaged over the region occupied by the Si-NCs. By obtaining $\omega(k)$ from the dispersion relation, we calculate the Purcell enhancement for the first, second, and third order modes as 3.8, 3.3, and 3.1, respectively. Another way to calculate the Purcell enhancement is as follows. We make the intuitive approximation that the standing wave modes in Figure \ref{fig:fdtd}(b)-(d) are modes of cavities spaced with the period of the SPP half wavelength, as those modes have the same $\lambda_{sp}$. We then calculate radiative quality factor ($Q_{rad}$) and the mode volume, defined as:
\begin{equation}
    V_{quant}= \frac{\int\!\!\!\int{\epsilon_{E}(x,z)|E(x,z)|^2dxdz}}{\textrm{max}\left[\epsilon_{E}(x,z)|E(x,z)|^2\right]}Y,
\end{equation}
the integral is taken over the unit cell of the SPP wave\cite{Hughes_PCmodeV}, and $Y$ is the length of the grating bars. Using the above figures, and noting that the quality factor is limited by the absorption factor $Q_{abs}\approx50$ (since $Q_{rad}\approx500\gg Q_{abs}$), we can calculate the average Purcell enhancement by using:
\begin{equation}
	F=\frac{3}{4\pi^2}\left(\frac{\lambda}{n}\right)^3\frac{Q}{V_{quant}}\overline{\left(\frac{E}{E_{max}}\right)^2},
\end{equation}
where the E-field intensity is again averaged over the region occupied by the Si-NCs. With $Y=10\mu$m as in the fabricated devices, we estimate the average Purcell enhancements to be 2.7, 2.7, and 2.4 for respectively, the first, second, and third order modes, which approximates the results from the previous method.

\begin{figure}[ht]
\centering
\includegraphics[width=3.3in]{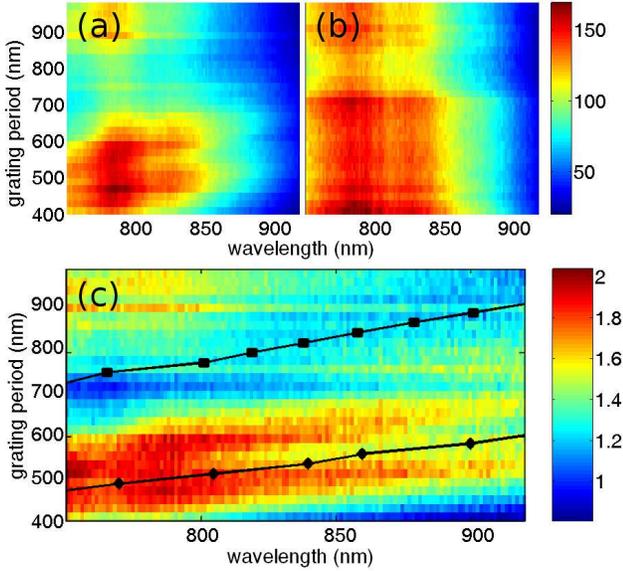}
\caption{PL from Si-NCs near the grating for the (a) p1 and (b) p2 polarizations. (c) The ratio of PL spectra for the p1 and p2 polarizations. The second (diamond) and third (square) order mode wavelengths from FDTD are plotted again from Figure \ref{fig:fdtd}(a).}
\label{fig:doublePL}
\end{figure}

The same designs are employed in our experiments. The experimentally observed enhancement in emission rate is:
\begin{equation}
    F_{meas}=\frac{\Gamma_{tot}}{\Gamma_{0}+\Gamma_{nr}} = \frac{\Gamma_{0}+\Gamma_{nr}+\Gamma_{pl}}{\Gamma_{0}+\Gamma_{nr}}
\label{Purcell_eq}
\end{equation}
where $\Gamma_{0}$ is the Si-NC spontaneous emission rate in bulk, $\Gamma_{pl}$ is the emission rate of the Si-NCs coupled to the plasmon mode, and $\Gamma_{nr}$ is the non-radiative recombination rate. While the measured enhancement differs from the calculated enhancement $F=\Gamma_{pl}/\Gamma_{0}$ from above, in the limit that the non-radiative decay rate is negligible ($\Gamma_{nr} \ll \Gamma_{0},\Gamma_{nr}$), the two figures are equal. On the same chip, we create gratings with a range of periodicities and analyze the PL for each grating period. The SPP grating modes in Figure \ref{fig:fdtd}(b)-(d) are predominantly polarized in the direction perpendicular to the grating bars (p1 in Figure \ref{fig:bulk}(c)). In order to separate the enhancement due to SPP modes from other effects, we isolate two polarizations (p1 and p2) in experiment by collecting with different polarizer settings. The results are shown in Figure \ref{fig:doublePL}(a)-(b). We observe that the PL polarized in the p1 direction (which coincides with the SPP polarization) has a shift in the wavelength with respect to the grating period, while the PL polarized in the p2 direction is fairly constant for all grating periods. The ratio of the two spectra, $PL_{p1}/PL_{p2}$, is then calculated and plotted in Figure \ref{fig:doublePL}(c). There is a noticeable shift in the peak of the PL enhancement, and this enhancement can be strictly attributed to the enhancement of coupling to the grating SPP modes. The band edge mode frequencies for the second order mode at the $\Gamma$ point and the third order mode at the $X$ point from Figure \ref{fig:fdtd}(a) are redrawn on top of the experimental data. The theoretical and experimental data are a good match, as we see enhancement for the second order mode, which is in the collection cone our objective, and suppression for the third order mode, which is outside of the collection cone. For the mode at the $\Gamma$ point, we obtain an experimental maximum Purcell enhancement of approximately 2. While such enhancements are comparable to previous references, we also note that the obtained Purcell enhancement is averaged over the exponential decay tail of the SPP mode. Peak enhancements for some nanocrystals are thus even larger, much like in the case where quantum wells of smaller thicknesses are placed at an optimal distance away from the metal-dielectric interface \cite{Scherer_SP}.

\begin{figure}[hb]
\centering
\includegraphics[width=3.3in]{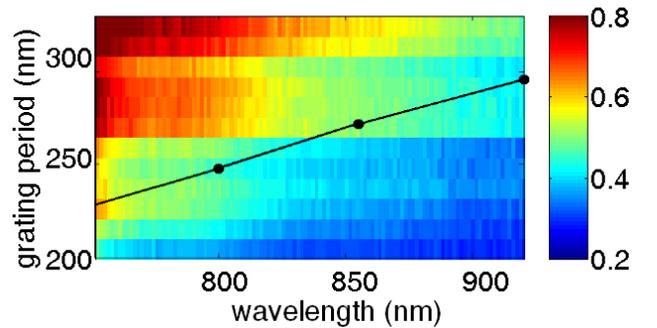}
\caption{$PL_{p1}/PL_{p2}$ from Si-NCs coupled to the first order SPP grating mode. The FDTD calculated first order mode wavelengths are plotted again from Figure \ref{fig:fdtd}(a).}
\label{fig:singlePL}
\end{figure}

To reach the range of the first order mode, we repeat the experiment for smaller grating periodicities (200-320nm) as well, and plot the same $PL_{p1}/PL_{p2}$ ratio in Figure \ref{fig:singlePL}. 
We find that the PL emission is suppressed for grating periods below 350nm by as much as a factor of 2, and the theoretically predicted first order mode wavelength matches the suppressed PL region in the middle and lower right of Figure \ref{fig:singlePL}. Again, because the first order grating modes fall below the light line, they will not be collected into the objective, leading to the suppression of emission. 
\begin{figure}[ht]
\centering
\includegraphics[width=3.3in]{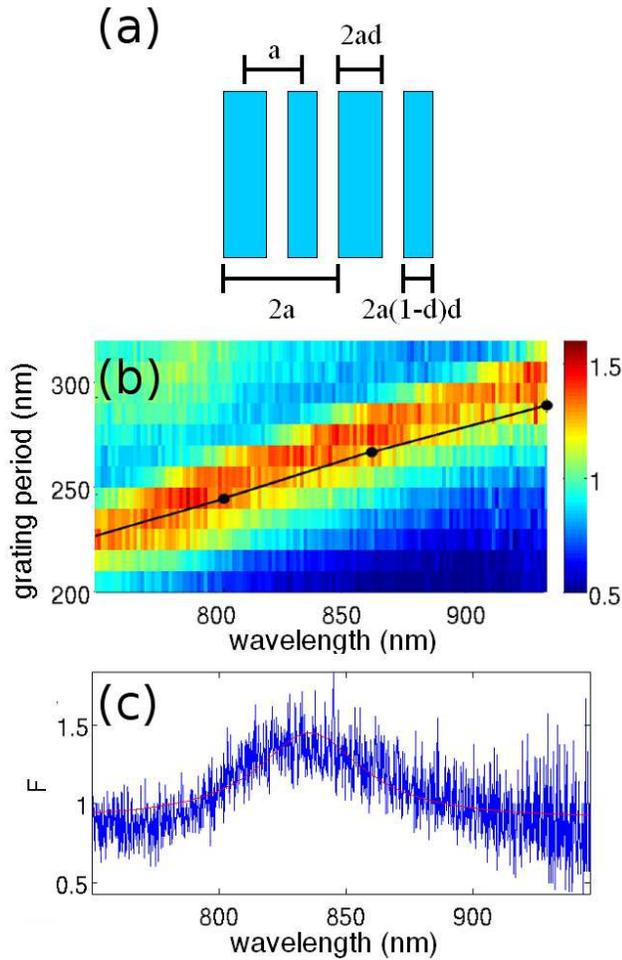}
\caption{(a) Biharmonic grating design, where $a$ is the grating period that couples Si-NC to the SPP mode at $\pi/a$, and $d$ is the duty cycle that can vary between 0 and 1. (b) The enhancement of the p1 polarization by the biharmonic gratings. The first order mode wavelengths from FDTD are plotted again from Figure \ref{fig:fdtd}(a). (c) Enhancement for one particular grating periodicity, with the red line being a Lorentzian fit of to data, representing $Q=16$.}
\label{fig:biharm}
\end{figure}

In order to extract the first order mode emission, we employ biharmonic gratings as discussed in previous works \cite{Hatano_biharm,Kocabas_biharm}. In all the examples above, the single periodicity of the gold bars, $a$, establishes the edge of the first Brillouin zone at the $X$ point. By introducing a secondary periodicity ($2a$) twice as large as the primary periodicity ($a$) into the gratings, we can scatter modes at the $k=2\pi/(2a)=\pi/a$ point back to the $\Gamma$ point, i.e., into the collection cone of the objective lens. In particular, the approach that was taken in this work was to create a skewed grating with a unit cell of two gold bars where the unit cell period is $2a$, the bar center-to-center width is $a$, the larger of the two bars has width $2ad$, and the ratio of the bar sizes is $1:(1-d)$ (Figure \ref{fig:biharm}(a)). The gratings with primary periodicities from 200-320nm and $d=0.3$ are fabricated with the above procedures and we plot the ratio $PL_{p1}/PL_{p2}$ once more for these structures in Figure \ref{fig:biharm}(b). We observe the same red-shift in the PL as the grating period increases, but also observe a much narrower linewidth than in the PL from the second order grating, corresponding to $Q=16$ from a fit to a Lorentzian (Figure \ref{fig:biharm}(c)). As before, we redraw the grating periodicity against band edges of the first order mode from the FDTD simulations above, and the peak enhancement follows the band edge at $\pi/a$ well, implying that the first order mode is successfully extracted via such biharmonic gratings.

In conclusion, we have demonstrated Purcell enhancement of Si-NCs coupled to SPP grating modes, with a maximum average enhancement of 2, closely matched by theory. By using a biharmonic grating, the first order SPP mode (at the $X$ point, which lies under the light line) is successfully extracted and re-radiated in the vertical direction, leading to similar enhancements with a $Q$-factor of 16. Such a device can be applied to an electrically driven LED, as the metallic portions of the device can also serve as contacts. In addition, coupling to SPP modes should also help reduce the lifetime of the Si-NCs, increasing the modulation rates for the device. Finally, by creating a metal-insulator-metal like structure for this device \cite{JV_sim}, it will be possible to achieve even higher Purcell enhancements, as well as simple electrical contacting.

The authors would like to acknowledge the MARCO Interconnect Focus Center and the NDSEG fellowship for funding, as well as Stanford Nanofabrication Facilities for fabrication facilities and Tom Carver for metal evaporations.

\end{document}